\pdfoutput=1 
\documentclass{JINST}

\title{First tests for an online treatment monitoring system with in-beam PET for proton therapy}

\author{A.C. Kraan$^{a*}$,
G. Battistoni$^b$, 
N. Belcari$^a$,
N. Camarlinghi$^a$,
F. Cappucci$^b$,
M. Ciocca$^c$,
A. Ferrari$^d$, 
S. Ferretti$^a$,
A. Mairani$^c$,
S. Molinelli$^c$,
M. Pullia$^c$,
A. Retico$^a$,
P. Sala$^b$, 
G. Sportelli$^a$, 
A. Del Guerra$^a$, 
V. Rosso$^a$\\
\llap{$^a$}Department of Physics, University of Pisa and INFN, Largo B. Pontecorvo 3, 56127 Pisa, Italy\\
\llap{$^b$}INFN Sezione di Milano, Via Celoria 16, 20133 Milano, Italy\\
\llap{$^c$}Fondazione CNAO, Strada Campeggi 53, 27100 Pavia, Italy\\
\llap{$^d$}CERN, Route de Meyrin 385, 1217 Meyrin, Switzerland\\
\llap{$^*$}E-mail: \email{aafke.kraan@pi.infn.it}}

\abstract{PET imaging is a non-invasive technique for particle range verification in proton therapy. It is based on measuring the $\beta^+$ annihilations caused by nuclear interactions of the protons in the patient. In this work we present  measurements for proton range verification in phantoms, performed at the CNAO particle therapy treatment center in Pavia, Italy, with our 10 x 10 cm$^2$ planar PET prototype DoPET. PMMA phantoms were irradiated with mono-energetic proton beams and clinical treatment plans, and PET data were acquired during and shortly after proton irradiation. We created 1-D profiles of the $\beta^+$ activity along the proton beam-axis, and evaluated the difference between the proximal rise and the distal fall-off position of the activity distribution. A good agreement with FLUKA Monte Carlo predictions was obtained. We also assessed the system response when the PMMA phantom contained an air cavity. The system was able to detect these cavities quickly after irradiation. 
}


\begin{document}

\section{Introduction}\label{sec:xxx}
Positron-Emission-Tomography (PET) is a non-invasive in-vivo method for monitoring the correctness of the dose delivery to patients undergoing particle therapy. During proton irradiation, various $\beta^+$ emitters (mostly $^{11}$C and $^{15}$O, etc) are produced,  whose subsequent decays into a photon pair can be recorded with a PET detector. Being based on different physics processes, no direct correlation exists between $\beta^+$- activity and dose. Nevertheless, by comparing pre-calculated Monte Carlo (MC) PET activity profiles with measured profiles, it is possible to verify that the proton range in the patient is as expected, assuring a correct dose delivery (see for reviews \cite{bib1, bib2, bib3}) 

Among the different strategies in PET imaging for proton range verification there are i) \emph{offline} data acquisition, where data are acquired with a commercial full-ring PET after patient irradiation outside the treatment room \cite{parodi_after, knopf, bauer}, ii) \emph{in-room} data acquisition, the same but with a full-ring PET detector installed in the treatment room \cite{zhu, min}, and iii) \emph{in-beam} data acquisition, where the PET system is integrated in the beam delivery system and data are acquired during or immediately after irradiation \cite{enghardt1, enghardt2, crespo, nishio, tashima} inside the treatment room. 

So far, the clinical use of PET imaging for proton range verification has been mostly focused on imaging after particle irradiation, providing retrospective information. However, it is recognized that acquiring PET data during irradiation has some important advantages with respect to imaging after irradiation~\cite{enghardt1, enghardt2}, among which the reduction of signal washout, smaller imaging times, sensitivity to intra-fractional errors, and the possibility for applying adaptive proton therapy~\cite{shao}. The latter becomes increasingly relevant when applying hypo-fractionational dose schemes, where precise dose monitoring is essential. The main difficulties of in-beam acquisition are beam-background during dose delivery and geometrical issues. Although the former can be overcome by applying a gating system so that data are acquired only in beam-pauses, geometrical issues remain a challenge. Indeed, no full-ring scanner can be installed and systems with partial angular coverage suffer from reconstruction artifacts~\cite{crespo2006}. Still, the advantages of PET imaging during irradiation are highly relevant for patients and couterbalance the disadvantages. 

PET data acquisition during particle irradiation has been discussed for instance in~\cite{enghardt1, enghardt2, crespo, sportelli}. Clinical studies on particle range verification in patients undergoing carbon therapy have been performed by Enghardt et al~\cite{bib3, enghardt1, enghardt2}, where data were acquired both after irradiation and during beam-pauses, i.e., in between the spills. Recently, Shao et. al. have presented a dual head PET prototype, which can provide information on particle range already during dose delivery, but it was tested on a very small scale (4.4 cm head distance)~\cite{shao}. In-beam PET imaging has also been discussed extensively elsewhere (see e.g. \cite{nishio}), but in these works data are presented, which are acquired immediately following irradiation and not during irradiation.

We have developed a compact planar PET system of 10x10 cm$^2$, which can acquire PET data during irradiation ('real-time'), as we demonstrated in the past at the CATANA cyclotron~\cite{sportelli, aafke}, where PMMA phantoms were irradiated with 62 MeV protons. In this study we present recent measurements for proton range monitoring performed with this system at the CNAO (Centro Nazionale di Adroterapia Oncologica) treatment centre in Pavia. As different from data presented in~\cite{rosso}, we now acquire data during proton irradiation and irradiate our phantoms not only with mono-energetic proton beams but also with clinical treatment plans of several Gy.

\section{Methods and Materials}\label{sec:xxx}
\subsection{PET system}
\begin{figure}
\begin{center}
\vspace*{-2cm}
\includegraphics[width=.7\textwidth]{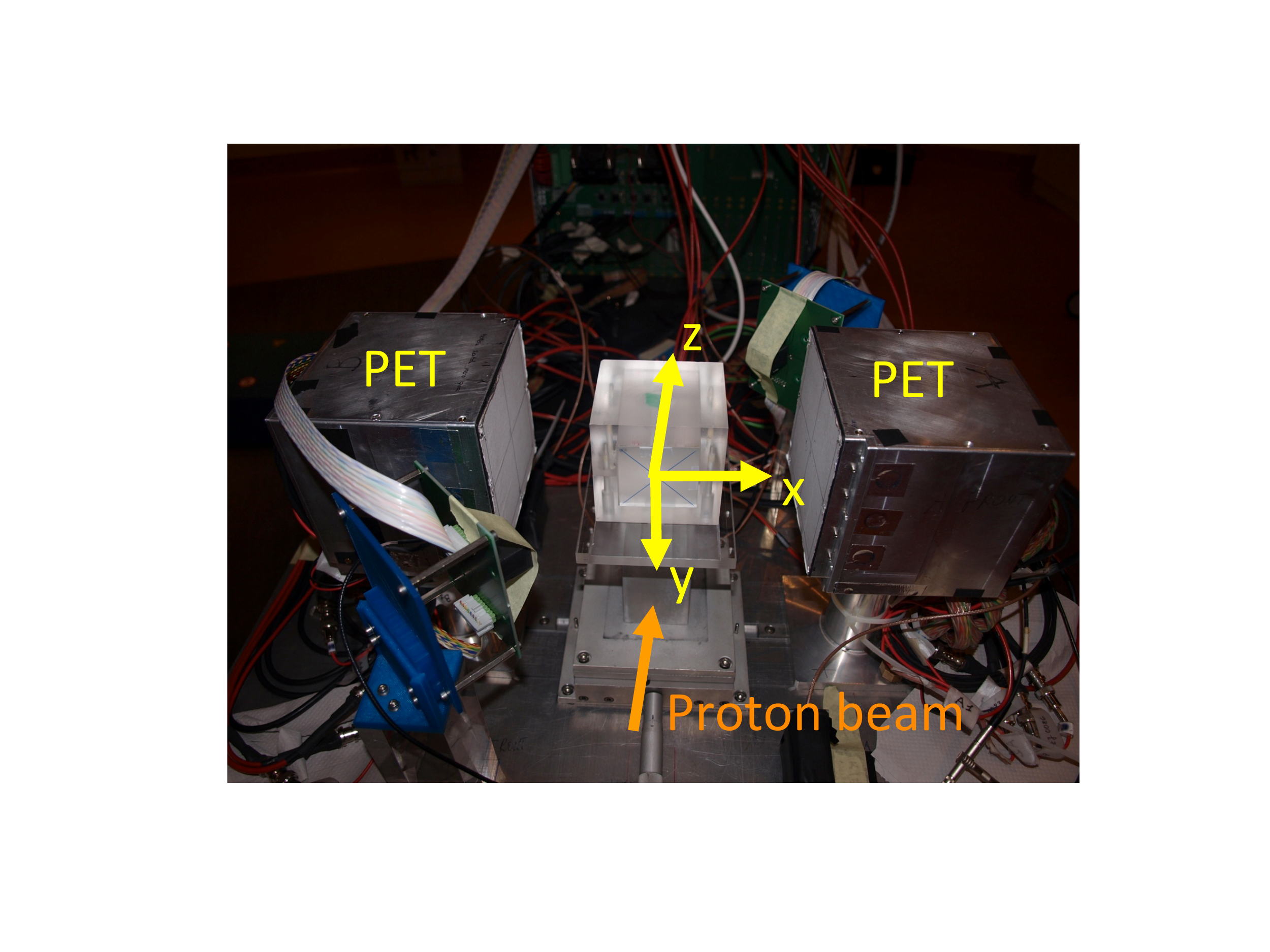}
\vspace*{-2cm}
\end{center}
\caption{Picture of experiment, with axes indicated. The proton beam is in the z-direction and enters perpendicular to the target. The y-axis points towards the ground. The x-axis is perpendicular to the PET planes. \label{figexperiment}}
\end{figure}

PET imaging was performed with an in-house developed system previously called DoPET, described in more detail in ~\cite{sportelli,aafke,rosso, attanasi, niccolo}. It was designed to be compatible with installation on a gantry for in-beam data acquisition. It was made of two planar 10 x 10 cm$^2$ detector heads placed 20 cm apart. Each head contained four modules of 5 x 5 cm$^2$. Each module was composed contained 23 x 23 LYSO crystals of 1.9 x 1.9 x 16 mm$^3$. The system utilized multi-anode position sensitive flat-panel photomultipliers and a fast front-end based on constant fraction discrimination. Data acquisition was performed by an FPGA. The coincidence window was 3 ns~\cite{sportelli}. The energy and spatial resolution was measured with a $^{22}$Na cubic source of 0.25 mm dimension. Due to the limited solid angle coverage the resolution of the detector was not uniform along x, y, and z. In the center of the Field Of View (FOV) the energy resolution (Full-Width-Half-Max, FWHM) was 18\% at 511 keV and the spatial resolution (FWHM) in x, y, and z was 7.0, 1.8 and 1.7 mm, respectively. The target was fixed inside a PMMA phantom holder, centred in between the two detector heads. The system was mounted with the detector planes parallel to the direction of the beam. In Fig.~\ref{figexperiment} we show the experimental setup and the reference system.
\subsection{Phantom irradiation at CNAO}\label{cnao}
Various homogeneous and inhomogeneous phantoms were irradiated with protons at the CNAO treatment centre~\cite{cnao} in Pavia, Italy. CNAO is a hospital-based facility, where since its opening in 2011 about 350 patients have been treated for various patologies, of which 100 with protons and the rest with carbon ions. Protons were extracted from a synchrotron with a spill-structure of 1 s beam, followed by a 4 s beam-pause. Phantoms were irradiated both with mono-energetic proton beams of 95 MeV (1.5$\times 10^{10}$ particles) and with a real treatment plan. The latter was made with the commercial Syngo VB10 RT Planning system (Siemens). Mono-energetic proton beams were impinging on the centre of the phantoms (see Fig 1). In the case of the treatment plans, irradiation was performed in sliced fractions ('iso-energy-layers') by means of spill-to-spill energy variation. For each energy layer the pencil beam moved thanks to fast scanning-magnets (modulated pencil beam scanning). The FWHM of the pencil beam depended on the beam energy and varied from 1.13 cm at 101 MeV to 1.51 cm at 50 MeV. The treatment plan delivered a dose of 2 Gy (total 2$\times 10^{10}$ particles) on a planning target volume (PTV) of  3x3x3 cm$^3$ located centrally at 3 to 6 cm depth. No ripple filter was used. Table 1 summarizes the irradiated phantoms, the irradiation time $t_{irr}$, and the number of reconstructed annihilation events for each acquisition. 
\begin{table}[b]
\small
\caption{Data acquisitions, phantoms and number of acquired coincidences.\label{table1}}
\begin{tabular}{|c|c|c|l|c|c|c|c|}
\hline
Acq. & Beam & Irradiation & Phantom & \multicolumn{4}{c|}{Number of acquired coincidences}\\
\cline{5-8}
& & time(s) & & In-spill & Inter-spill & Beam-off & Total\\ 
\hline
\vspace*{-0.4cm}
 & & & & & & & \\
1 & E=95 MeV & 118 s & PMMA \includegraphics[width=0.09\textwidth]{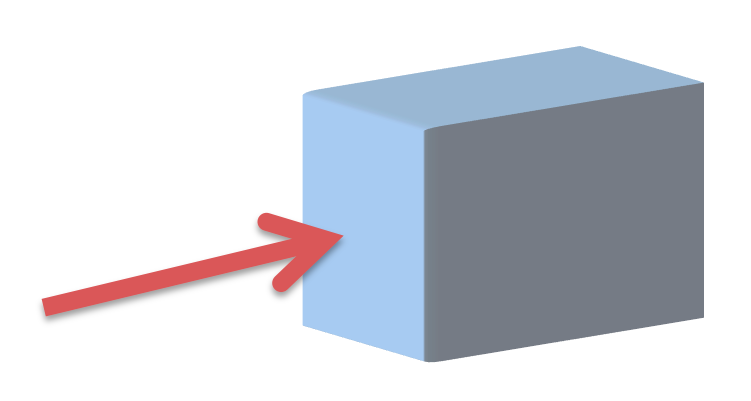} & 1.2E3 & 2.0E5 & 3.3E5 &5.3E5\\
\vspace*{-0.5cm}
 & & & & & & & \\
2 & E=95 MeV & 90 s & PMMA+cavity 1 \includegraphics[width=0.09\textwidth]{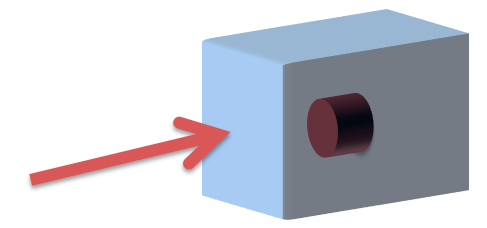} & 9.1E3 & 1.8E5 & 3.5E5 & 5.4E5\\
\vspace*{-0.5cm}
 & & & & & & & \\
\hline
\vspace*{-0.4cm}
 & & & & & & & \\
3 & 2 Gy & 145 s & PMMA \includegraphics[width=0.09\textwidth]{PMMA.png} & 9.3E3 & 1.8E5 & 2.8E5 & 4.7E5\\
\vspace*{-0.5cm}
 & & & & & & & \\
4 & 2 Gy & 160 s & PMMA+cavity 2 \includegraphics[width=0.09\textwidth]{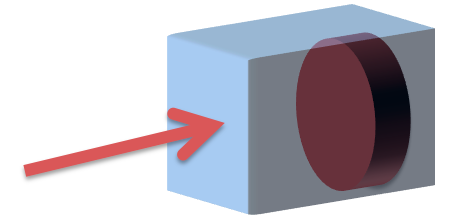} & 1.1E4 & 2.0E5 & 2.8E5 & 4.9E5 \\
\vspace*{-0.5cm}
 & & & & & & & \\
5 & 2 Gy & 286 s & PMMA \includegraphics[width=0.09\textwidth]{PMMA.png}& 9.0E4  & 3.4E5 & 2.2E5 & 6.5E5 \\
\hline
\end{tabular}
\end{table}
Schematic displays of the phantoms are also shown. 

Acquisition 2 and 4 are done with phantoms containing a cilindrical air cavity (cilinder axis=z-axis). For acquisition 2 the cilinder diameter and height was 1 cm, and it was located from z=1 to z=2 cm ('cavity 1'). For acquisition 4 the cilinder diameter was 3.4 cm and height 0.5 cm, and it was located from z=4 to z=4.5 cm ('cavity 2'). Although acquisition 5 was not used further in the data analysis, we include the statistics in Table 1, because the number of in-spill events in acquisition 5 is much larger than in the other data acquisitions. Interestingly, in this acquisition the dose rate was lower than in the other acquisitions, possibly causing the system to be less paralyzed as in the other acquisitions (see also section 4). 

PET data were acquired during irradiation, i.e., during and in between spills, and until 15 minutes afterwards. Since it is clinically highly relevant to get quick feedback from the system, the results used below are based on data until 2 minutes after irradiation. This would imply only a small increase in treatment room occupancy and patient discomfort, as well as a minimum amount of signal washout. 

\subsection{Image reconstruction}\label{sec_dataprocessing}
 The recently implemented possibility to acquire the signal from the synchrotron allowed us to tag the data as acquired during spills ('\emph{in-spill}'), in the beam-pauses between the spills ('\emph{inter-spill}'), and 2 minutes after data taking when there is no beam ('\emph{beam-off}'). When all acquired coincidences up to 2 minutes after irradiation are included we refer to it as '\emph{all data}'. PET data were acquired in the form of photons pairs in an energy window of [350, 850] keV. Image reconstruction was performed with the iterative Maximum Likelihood Expectation Maximization (MLEM) algorithm. For more information on the MLEM method and on image reconstruction techniques in PET the reader is referred to reference~\cite{reco}. A direct normalization technique was applied to account for issues such as non-uniform detector efficiency, whereby a planar source filled with FDG was placed in between the two detector prior to data taking. A standard correction for random coincidences was also included. No attenuation correction was performed, which is not expected to influence the fall-off position of the measured activity profile, as shown in ~\cite{aafke}. The total reconstructed field-of-view (FOV) was 10 x 10 x 10 cm$^3$ and each voxel was 1 x 1 x 1 mm$^3$. More details about the image reconstruction employed here are given in ~\cite{niccolo}.  

\subsection{Monte Carlo simulations}
\begin{figure}[b!]
\centering
\includegraphics[width=.4\textwidth]{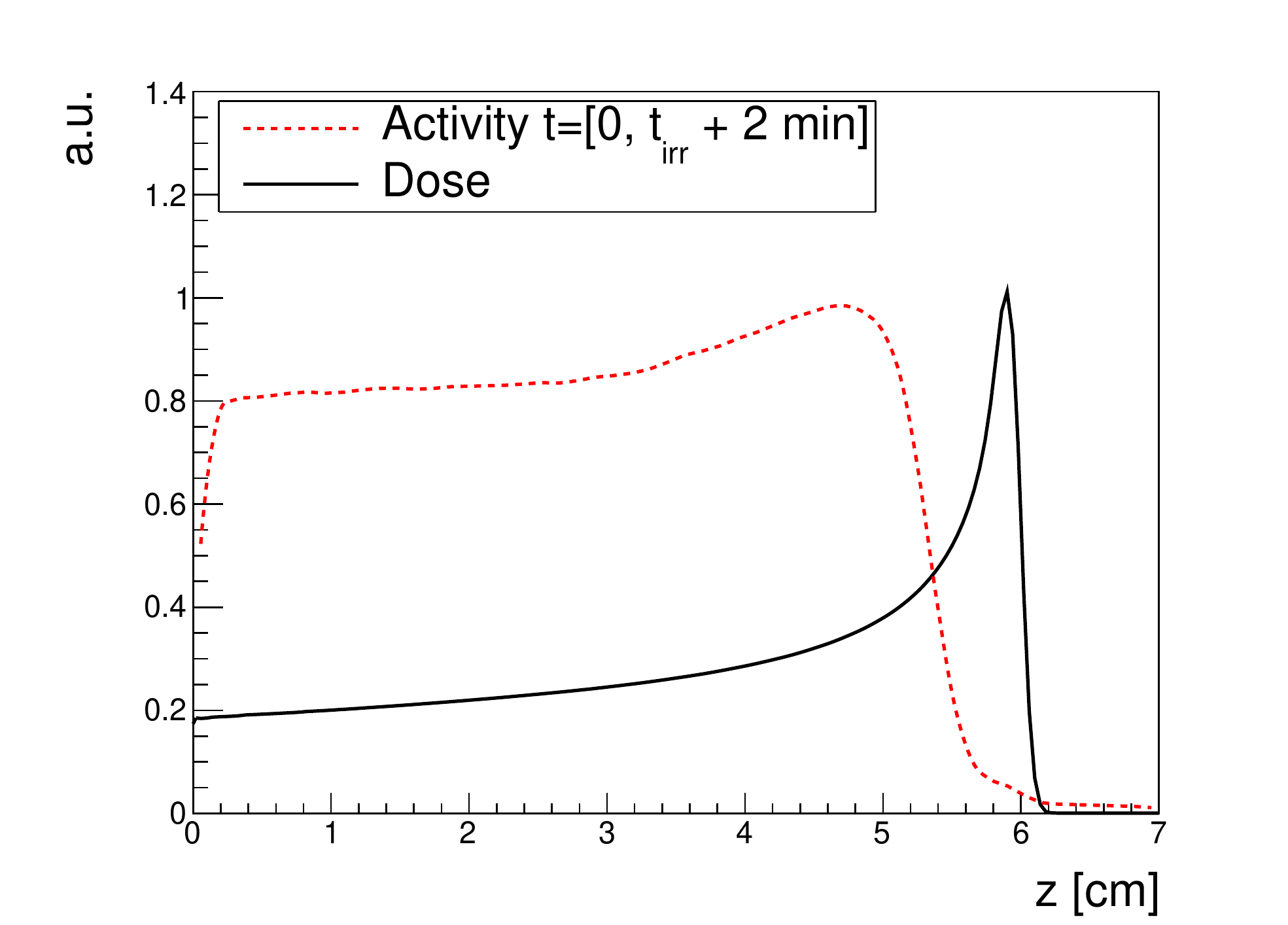}
\includegraphics[width=.4\textwidth]{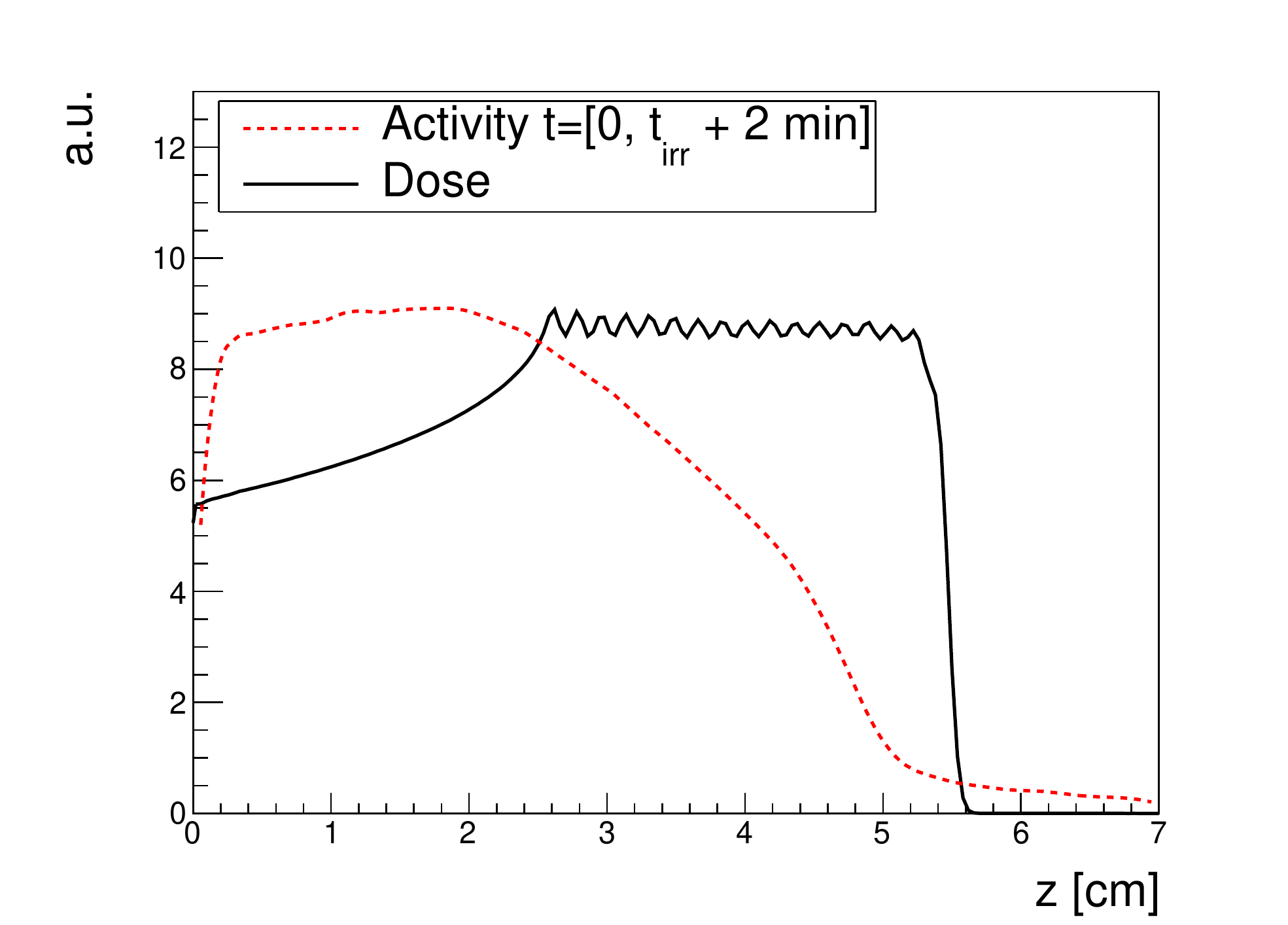}
\caption{1-D activity profile together with the dose distribution obtained from MC simulations for a 95 MeV proton beam (left) and a 2 Gy treatment plan (right). Units are arbitrary. The dose profile of the 2 Gy treatment plan is somewhat rippled because no ripple filter was used.}
\label{fig_dose}
\end{figure} 
The FLUKA code~\cite{fluka1,fluka2} was used to simulate proton interactions, where a user interface was developed to record the $\beta^+$-activity and annihilation products in space and time, as described in ~\cite{aafke}. We used Fluka2013 Version 0.0 (5/16/2013), including updated experimental nuclear interaction cross section data taken from the available data from the OECD (Organisation for Economic Cooperation and Development) Nuclear Energy Agency Data Bank. We used a previously developed in-house code~\cite{mairani} for including the CNAO beamline geometry and for converting the treatment plan (DICOM) into a text file which could be read in by FLUKA. Furthermore the phantom geometry was included in FLUKA. To account for the finite detector resolution, smearing with a Gaussian distribution was applied, where the $^{22}$Na measurements were used to determine the smearing parameters. For each data acquisition we simulated the corresponding MC sample containing 200M initial protons. Figure~\ref{fig_dose} shows the FLUKA MC simulated dose and activity for a 95 MeV proton beam (left) and the 2-Gy treatment plan (right) impinging on a PMMA target.  
Finally, it must be noted that the MC distributions shown are pure MC truth distributions with photons emitted in a 4$\pi$ geometry, and that no full photon propagation and simulation of the detector response and signal acquisition was included in the simulation (see discussion in section 4). 

\subsection{Data analysis\label{analysis}}
\begin{figure}[h]
\begin{center}
\hspace*{-1.5cm}\includegraphics[width=1.2\textwidth]{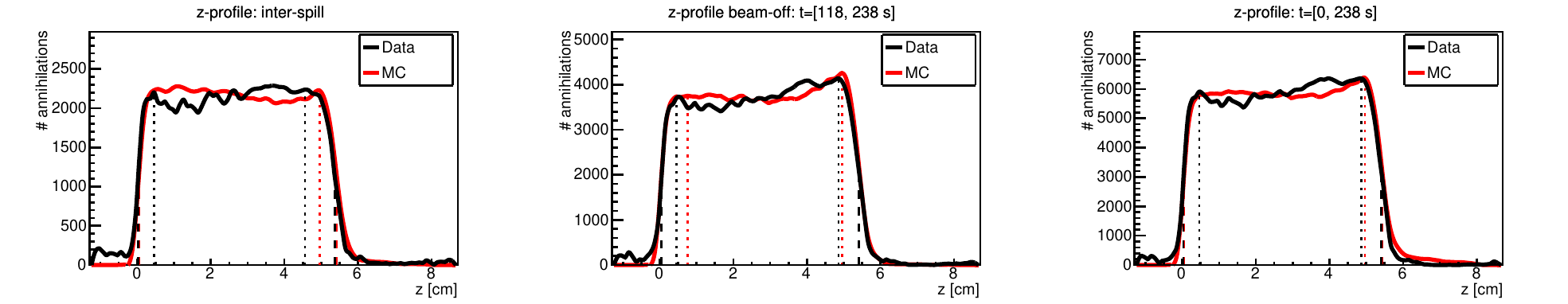}\\
\hspace*{-1.5cm}\includegraphics[width=1.2\textwidth]{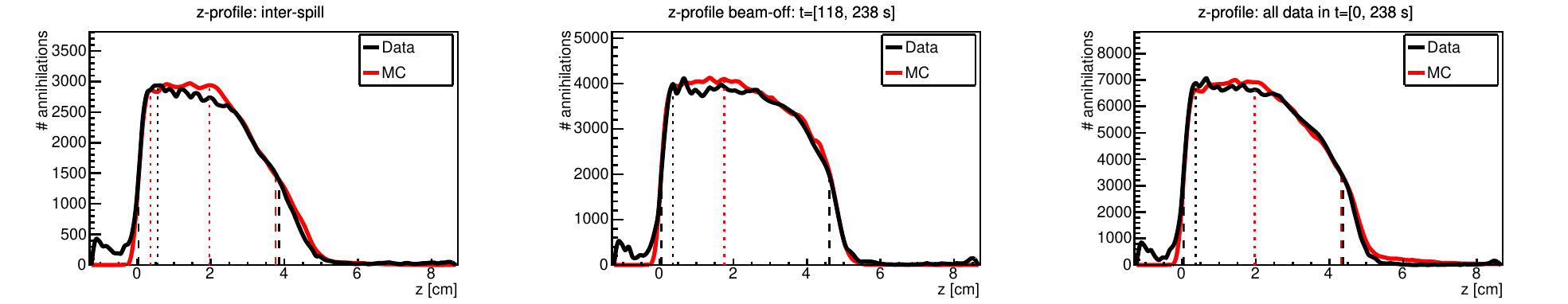}
\end{center}
\caption{Top: 1-D profile along the z-axis for a 95 MeV proton beam (acquisition \#1) of the measured (black) and MC simulated activity (red), whit the determination of $\Delta w_{50\%}$. The black and red dotted lines indicate the maxima for data and MC, respectively. The dashed lines indicate the 50\% values. Where the maximum is located in the same bin, only one dotted line is visible. Bottom: the same but for the 2 Gy treatment plan (acquisition \#3). \label{fig_evaluation}}
\end{figure}

The current analysis focuses on two issues. First of all, we investigate the ability of our system to verify particle range in homogeneous PMMA phantoms, both for a 95 MeV proton beam (acquisition 1) and for a 2 Gy treatment plan (acquisition 3). For this purpose we apply a  commonly utilized procedure~\cite{bib2, parodi_after, enghardt1, shao, sportelli}, consisting of constructing 1-D activity profiles along the beam-axis, and comparing distal fall-off positions at 20\% and 50\% of the last local maximum in the depth-activity profiles between PET measurements and FLUKA MCpredictions. First the 1-D activity profile was made from the 3-D distribution by projecting all annihilation events that were present in a certain Region-of-Interest (ROI), here an elliptic cilinder with the beam-axis as central axis, similar as done in ~\cite{sportelli}. Then we evaluated the activity profile depth at 20\% and 50\% of the local maximum of the rise and fall-off, as demonstrated in Fig.~\ref{fig_evaluation} for the mono-energetic beam (top figures) and the treatment plan (bottom figures). 
For the mono-energetic beam, it is clear which maximum is related to the rise and fall-off of the signal. For the treatment plan however, it's questionable where to define the maximum of the signal fall-off. We stick to the same procedure as used for the mono-energetic beam, but we determine the position of the maximum of the signal fall-off from the MC simulation. The activity profile depth at 50\% of the maximum, $\Delta w_{50\%}$, was defined as the horizontal difference between the 50\% proximal rise and the 50\% distal fall-off of the 1-D profile along the beam-axis (z-axis), and analogously for the depth at 20\%, $\Delta w_{20\%}$. Linear interpolation between the data points was applied for finding precisely the values for the rise and fall-off position at a given percentage. 
We emphasis that our quantities $\Delta w_{50\%}$ and $\Delta w_{20\%}$ are just  measures to facilitate the comparison between data and MC simulations, but that other quantities or procedures may be equally valid for this purpose. 

Second, we investigate the capability of the system to quickly detect small modifications.  This was done by irradiating phantoms containing air cavities, with varying position and dimension as described in section~\ref{cnao}. A new activity profile was created, but with smaller ROI to emphasis the cavities, and compared with the default phantom. This study was performed for the mono-energetic proton beam (acquisition \#1 versus \#2) and a 2 Gy treatment plan (acquisition \#3 versus \#4).

\section{Results}
\subsection{Measurements for range verification}



In Figure~\ref{fig_evaluation} we show the measured and simulated activity profile for a 95 MeV proton beam (top figures) and a 2 Gy treatment plan (bottom figures) impinging on a PMMA target. For each we separately display the inter-spill acquisition (left), beam-off acquisition (middle) and all data together (right). It is also illustrated how we determine the rise minus fall-off position for data and Monte Carlo. The agreement between MC and data is satisfactory. Statistics for the in-spill case was not sufficient for making a valuable z-profile (see discussion). 
In Table~\ref{table2} we report the values for the 50\% and 20\% distal fall-off position of the 95 MeV proton beam and the 2 Gy plan on PMMA, for inter-spill acquisition, beam-off (2 minutes) acquisition and both together. Errors on these numbers are discussed below in section~\ref{errors}. 
\begin{table}[bp]
\small
\caption{\label{table2}Data acquisitions and values for $\Delta w_{50\%}$ and $\Delta w_{20\%}$ in MC and data. Errors on the numbers are discussed in section 3.2.}          
\begin{tabular}{|c|c|c|c|c|c|c|c|}
\hline  
Acquisition & Beam & Phantom & Data tag & \multicolumn{2}{c|}{$\Delta w_{50\%}$ (mm)} & \multicolumn{2}{c|}{$\Delta w_{20\%}$ (mm)}\\
\cline{5-8}
 &  & & & MC & Data & MC & Data \\
\hline
\hline
& &                & Inter-spill            & 53.8 & 53.5 & 57.1 & 56.7 \\
1 & 95 MeV protons & PMMA &Beam-off (2 min) & 53.8 & 53.7 & 56.9 & 57.0 \\
& &                & All                    & 53.9 & 53.8 & 57.2 & 56.9 \\
\hline
 & & &                      Inter-spill & 37.2 & 38.2 & 46.6 & 45.8  \\
3 & 2 Gy plan & PMMA & Beam-off (2 min) & 45.7 & 45.7 & 50.0 & 50.1 \\
 & & &                              All & 42.8 & 43.4 & 49.4 & 49.2  \\
\hline
\end{tabular}
\end{table}

\begin{figure}[t]
\begin{center}
\includegraphics[width=.6\textwidth]{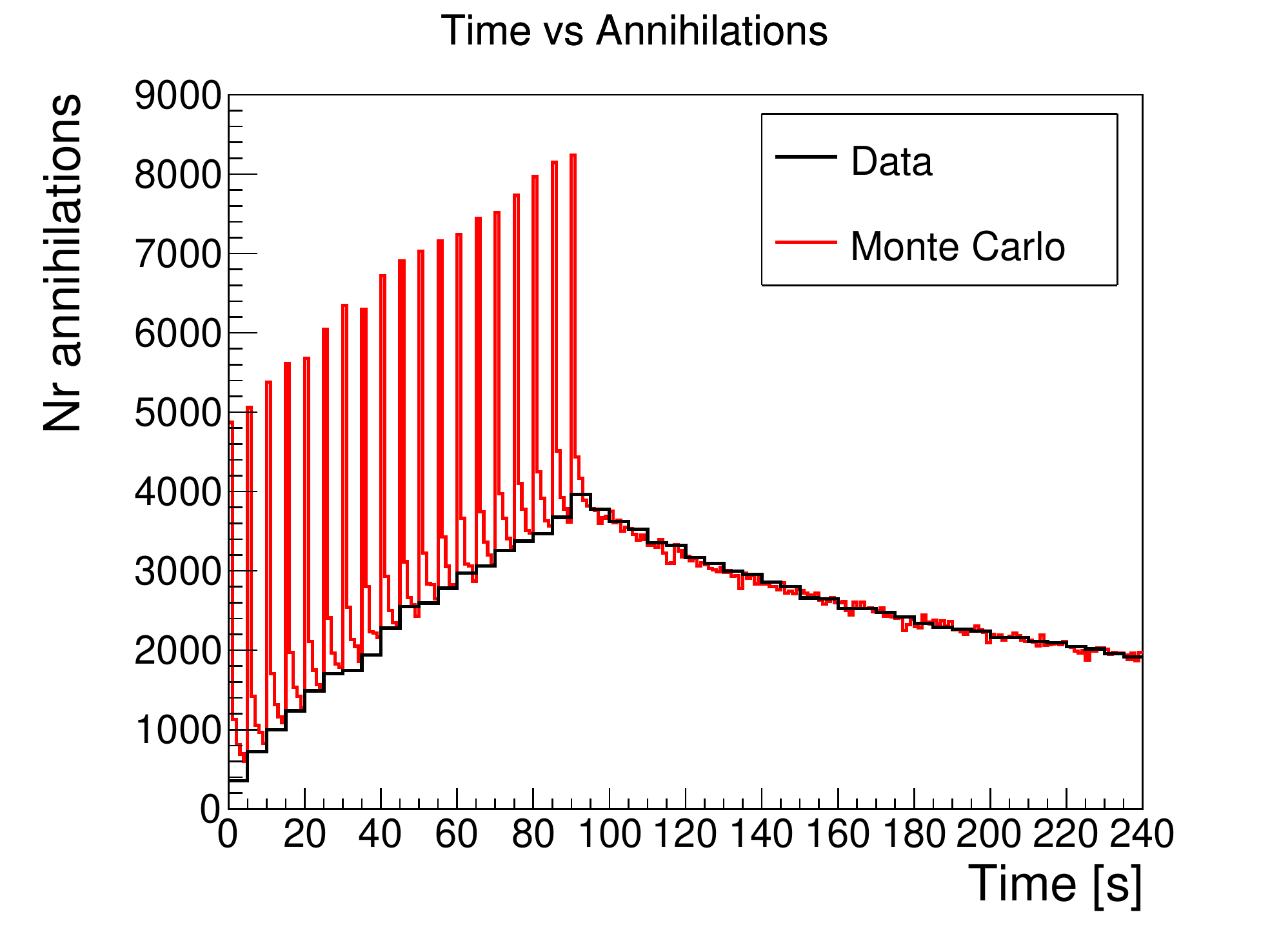}
\end{center}\caption{Measured time profile (black) together with the MC expectation (red) of one of the data acquisitions (nr 2). \label{fig_time}} 
\end{figure}

%
Figure~\ref{fig_time} shows an example of the number of annihilations versus time for data (black) and MC simulations (red).  

\subsection{Evaluation of uncertainties}\label{errors}
\begin{figure}[bp]
\includegraphics[width=.4\textwidth]{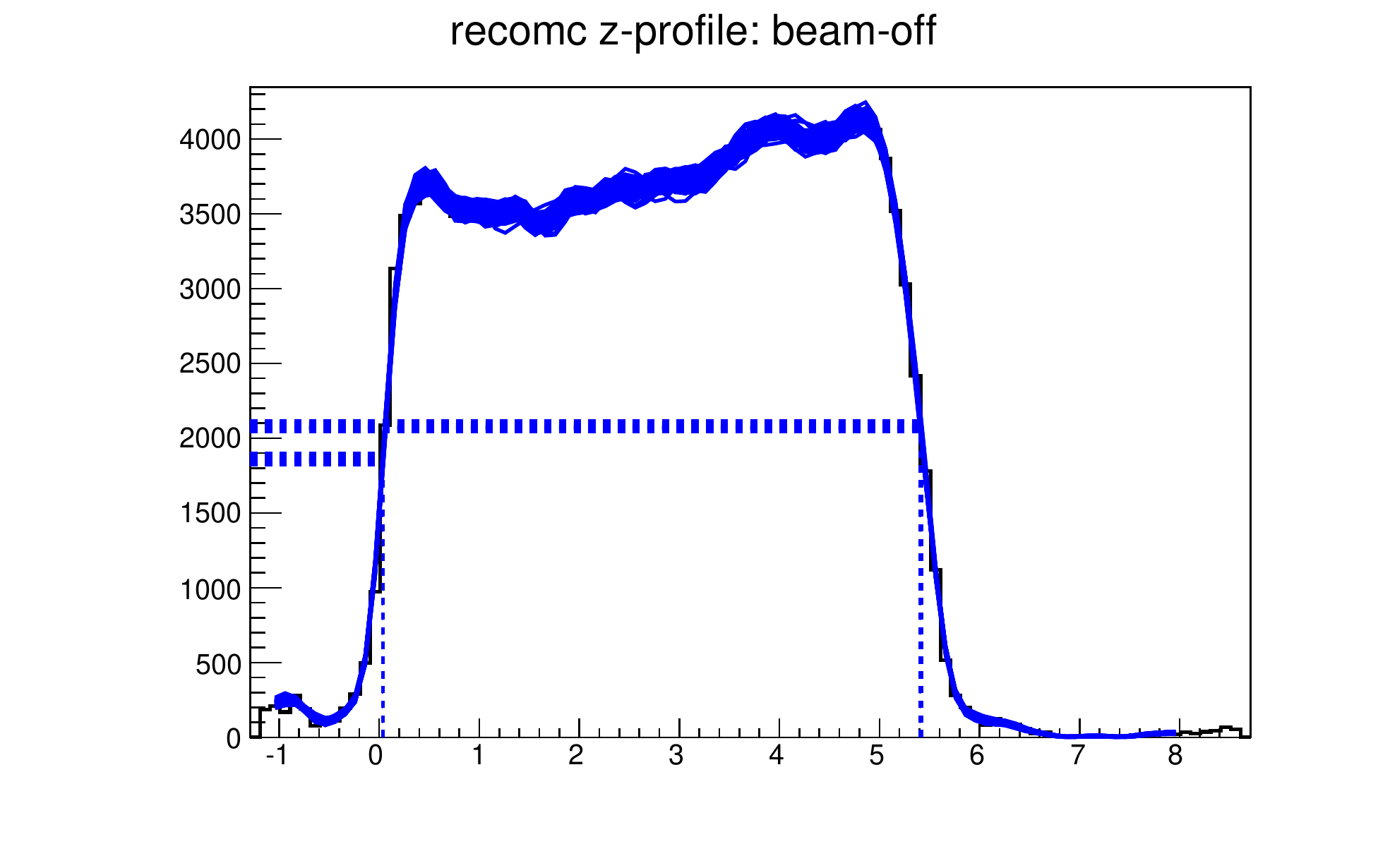}
\includegraphics[width=.4\textwidth]{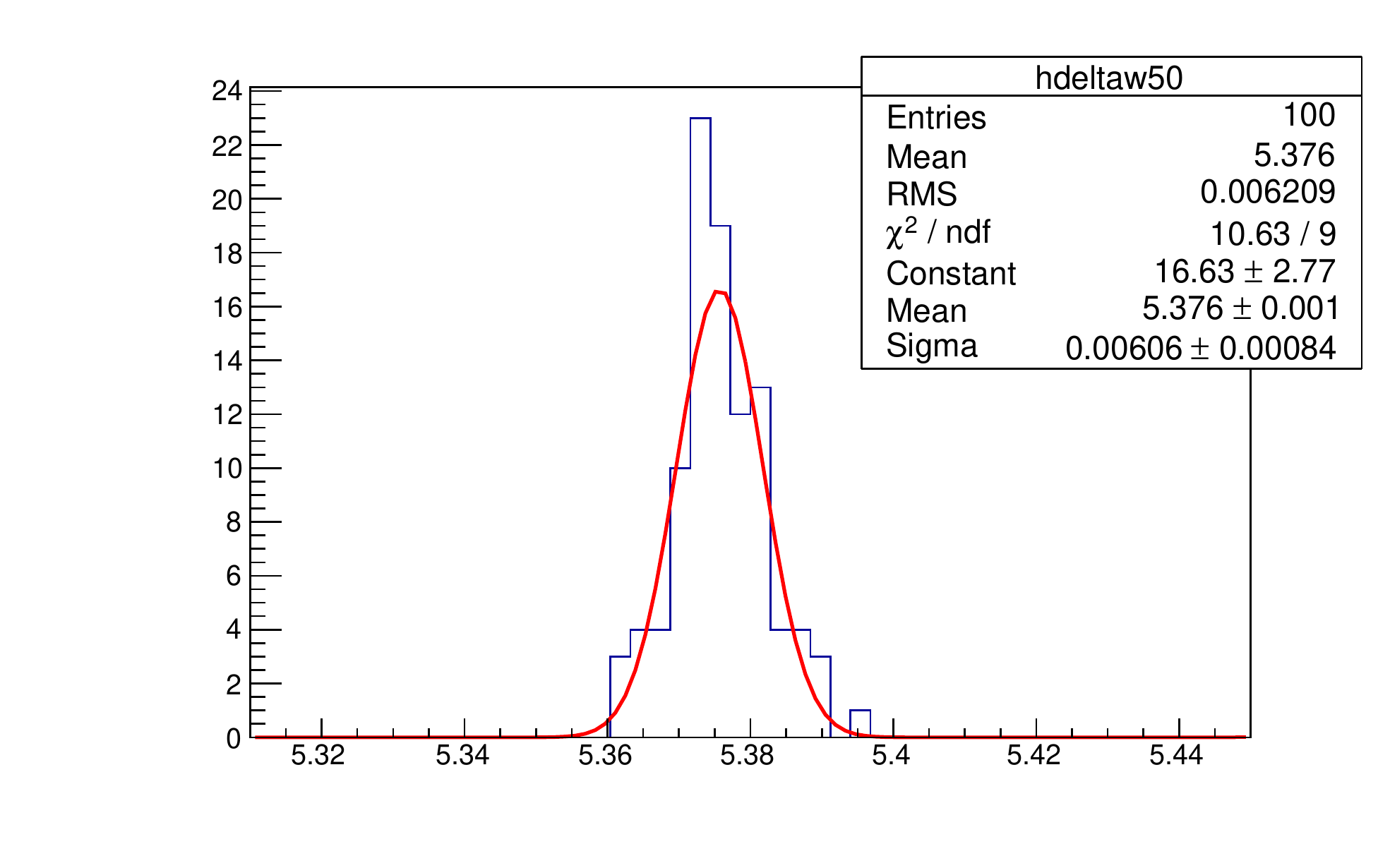}
\caption{Left: 1-D activity profiles for 100 pseudo-experiments (blue), with the values for the 50\% rise and fall-off indicated. Right: distribution of the obtained values for $\Delta w_{50\%}$ in the 100 pseudo-experiments. \label{fig_errors}} 
\end{figure}

There are various uncertainties which can influence our data profiles and the values reported in Table 2. The sensitivity of the procedure to statistical fluctuations was evaluated by performing pseudo-experiments, varying the bin content of each bin $i$ of the measured data profiles according to a Gaussian distribution with $\sigma_i$=$\sqrt{N_i}$, with $N_i$ the measured number of annihilation events in bin $i$. The procedure is displayed in Fig.~\ref{fig_errors} (left) for the 95 MeV proton beam (beam-off). In each pseudo-experiment the value for $\Delta w_{50\%}$ is determined. The obtained value for $\Delta w_{50\%}$ in the pseudo-experiments is fitted with a Gaussian, as demonstrated in Fig.~\ref{fig_errors} (right). The value of the statistical error is the $\sigma$ of the gaussian and was roughly 0.2\%, 0.1\% and 0.1\% for $\Delta w_{50\%}$ in inter-spill data, beam-off data and all data. The same procedure was applied to the treatment plan, resulting in a statistical error of about 0.5\% in all cases. 

Systematic uncertainties dominate the measurements. We have investigated various sources of uncertainties. For instance, in the reconstruction algorithm we tested the effect of using a different planar source for correcting the uniform detector response. The error in $\Delta w_{50\%}$ and $\Delta w_{20\%}$ was both for the treatment plan and for the mono-energetic beam about 1\% for the various time intervals. Other uncertainties may arise from the reconstruction algorithm including the calibration procedure, the MLEM iterations, histogram binning and voxilization, and so on. As discussed in Section 4, a detailed analysis of such uncertainties is beyond the scope of the current work.

For the MC simulations, we only evaluated statistical uncertainties, and errors were found to be similar to what was found for data. Systematic uncertainties dominate here too. Seravalli et. al. evaluated the difference in range for 5 different MCcodes including FLUKA, showing a variation in particle range of 2\% for 90 MeV protons in PMMA between MC codes (see table 2 in ~\cite{seravalli}). 
 
\subsection{Detecting anomalies}
\begin{figure}[tbp]
\includegraphics[width=.4\textwidth]{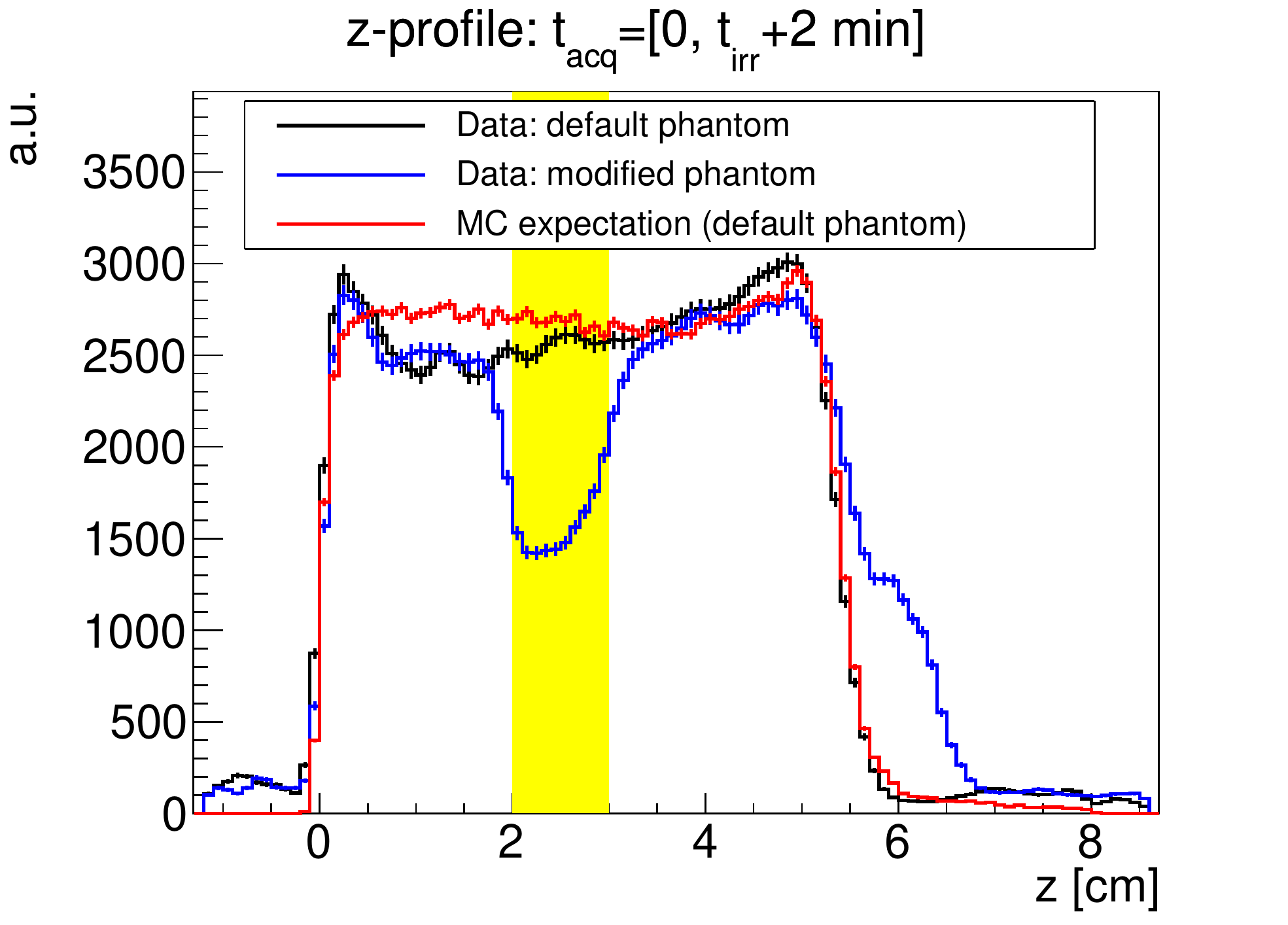}
\includegraphics[width=.4\textwidth]{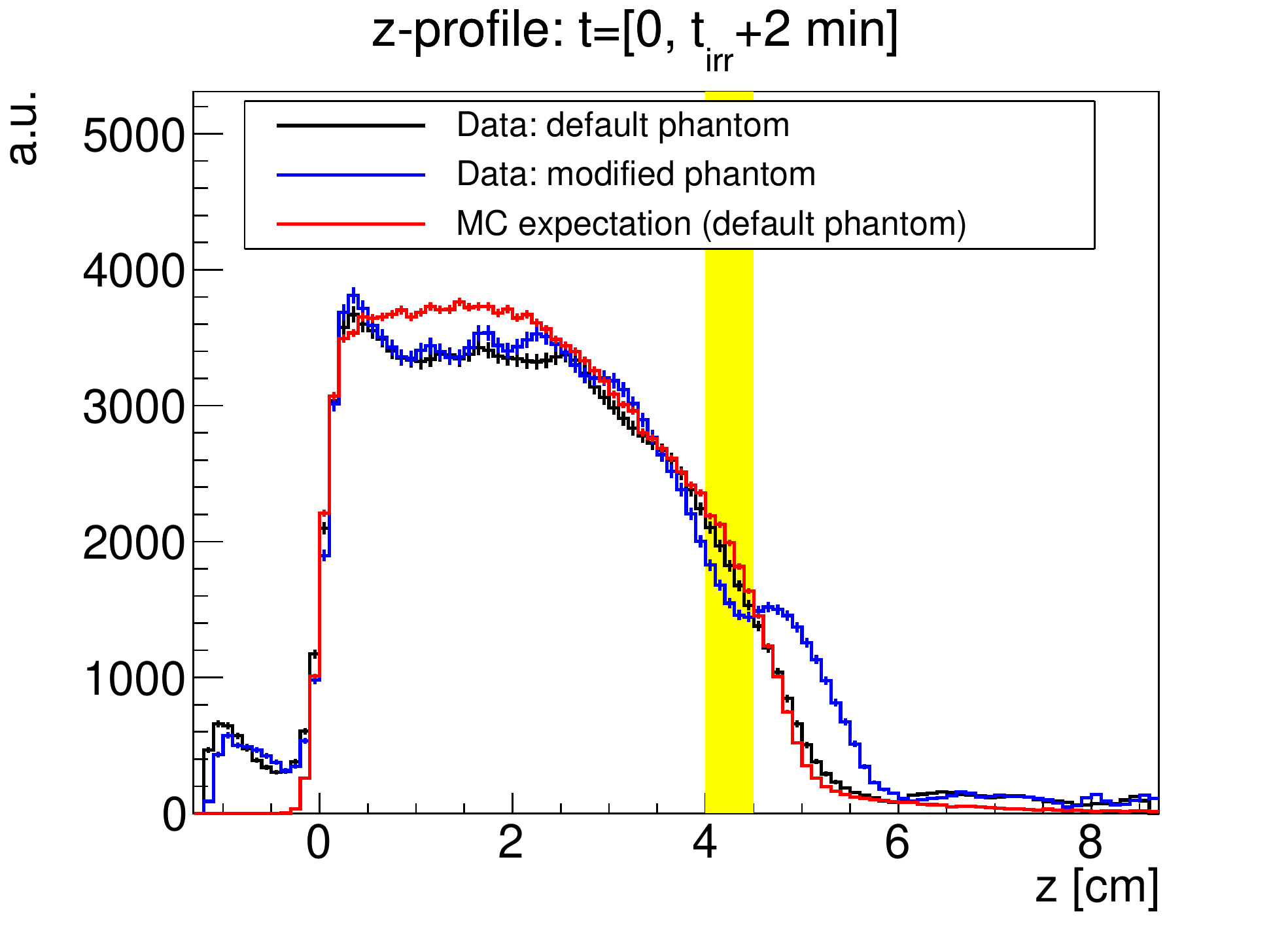}
\caption{Left: 1-D activity profiles along the beam direction for an irradiation of PMMA with 95 MeV protons: data for a default homogeneous PMMA phantom (red), FLUKA MC simulation (blue) and data for a PMMA phantom including a cavity at 2 cm depth ('modified').  (b) The same, but for a 2 Gy plan: data for default case (red), MC expectation (blue), and data for a phantom with a cavity at 4 cm (black). The yellow band indicates the z-position of the cavities for the modified phantom.\label{fig_MCdata}} 
\end{figure}

In fig.~\ref{fig_MCdata} we show an example of how an anomalous situation could appear in our data profiles. On the left side of fig.~\ref{fig_MCdata} we show the expected (red line) and measured (black line) profile of a mono-energetic proton beam on PMMA ('default phantom'). In the same figure is shown how the profile changes when the PMMA target contains a small central cilindrical air cavity of diameter d=1 cm and height h=1 cm (blue line), located centrally ('modified phantom'). On the right side of fig.~\ref{fig_MCdata} the same is demonstrated for the 2 Gy treatment plan (fig.~\ref{fig_MCdata} right), where the modified phantom contained a cavity (d=3.4 cm, h=0.5 cm) located at 4 cm depth. In fig.~\ref{fig_MCdata} we included all data acquired during irradiation until 2 minutes after irradiation, and we included a smaller ROI than in Fig.~\ref{fig_evaluation} to evidence better the hole. 

\section{Discussion}
In the current study we demonstrated the ability of the system to give valuable feedback on particle range in homogeneous PMMA phantoms within 2 minutes after irradiation. The activity profiles in Fig.~\ref{fig_evaluation} and the values for $\Delta w_{20\%}$ and $\Delta w_{50\%}$ reported in Table 2 showed that the measurements agree well with the predictions from the FLUKA MC generator for various time intervals. Some differences in profile shape were observed. As discussed also in~\cite{aafke}, it is likely that these differences were caused by the absence of a detailed MC simulation framework, including photon propagation through the detector, electronics, and reconstruction. Such a framework is also required for fully assessing the effect of uncertainties on the range verification measurements and MC predictions. We are currently working on such a detailed description. Despite this drawback, it is possible to predict the activity depth with MC and we were able to address the main question here whether our system was able to give reliable feedback on the particle range at CNAO under clinical conditions. 

Table 2 showed that the values for $\Delta w_{20\%}$ and $\Delta w_{50\%}$ for the inter-spill measurements of the treatment plan irradiation are smaller than the values obtained for beam-off acquisition. For the mono-energetic beam, values were instead similar. This is explained by the time-course of the irradiation. While for a mono-energetic beam, the beam was shot on one position continuously, for the treatment plan, the x, y, and z-position of the beam was changing over time. At CNAO, dose delivery to the target with a treatment plan starts by irradiating the shallowest energy layer, and goes to deeper energy layer, finishing at the deepest energy layer. Acquiring PET data means detecting decays from nuclei produced at an earlier time in the target.  Therefore inter-spill data contain relatively more annihilations from irradiation of shallower energy layers, resulting in a more narrow activity profile width. Instead the beam-off data contain also activity generated in the last layers of the treatment plan, resulting in a wider profile. For mono-energetic beams this effect is absent. A detailed study on the time course of the irradiations is ongoing.

In-spill data taking remains a challenging issue. Looking at Table 1, we note that the in-spill contribution for acquisitions 1 to 4 was very small compared to the inter-spill contribution. This is also confirmed by the time profile shown in fig.~\ref{fig_time}, where we see that the activity produced within the spills is actually not present in data (even when decreasing bin size). The small amount of in-spill data is explained by the extremely high counting rate present during spills, causing the system to partially paralyze. Nevertheless the system is able to acquire valid data during the irradiation and to follow the increase  in counts like in the MCsimulations. Acquisition at lower dose rates probably makes in-spill data taking easier. In fact, Table~\ref{table1} shows that, when the beam intenstity is reduced, like in acquisition \#5, a significant amount of in-spill annihilations were acquired. More experimental data at lower dose rates are needed to understand more precisely how to profit from the in-spill data. 

When introducing small cavities in the PMMA phantoms, the system response was very clearly different as it was expected. This makes us confident that this prototype can be a starting point for an in-beam PET treatment monitoring system at CNAO. A larger size system would help in improving accuracy, and in extending the reach in depth of the system. We are currently upgrading our detector to 15x15 cm$^2$, and in addition there is a Time-Of-Flight PET system under construction of 10x25 cm$^2$ (INSIDE project).

\section{Conclusion}
Real-time treatment monitoring is attractive, avoiding signal washout and providing immediate information about the treatment.
First tests on PMMA phantoms for using our PET system as real-time monitoring device have been performed at the CNAO treatment facility in clinical conditions. The agreement between FLUKA MC predictions and CNAO data was satisfactory for all cases tested. First tests with phantoms containing air cavities show that our PET system is able to detect anomalies in PMMA within a few minutes after the end of irradiation. 

\section{Acknowledgements}
This research has been supported by the FULLBEAM-300838 Marie Curie Intra European Fellowship within the 7th European Community Framework Programme, and by the INFN- RT 60141 POR CRO FSE 2007-2013 fellowship.
%
%
%
%
%
%
%
%
%
%
%
%

\end{document}